# Lunar Dust Particles Blown By Lander Engine Exhaust in Rarefied and Compressible Flow

J. E. Lane[1], P.T. Metzger[2], J.W. Carlson[3]

[1] ASRC Aerospace, Kennedy Space Center, FL  32899, John.E.Lane@nasa.gov
[2] Granular Mechanics and Regolith Operations Lab, NASA, Kennedy Space Center, FL 32899, Philip.T.Metzger@nasa.gov
[3] ASRC Aerospace, Kennedy Space Center, FL  32899, Jeffrey.W.Carlson@nasa.gov

**ABSTRACT**

At Earth & Space 2008, we presented a numerical model that predicts trajectories of lunar dust, soil, and gravel blown by the engine exhaust of a lunar lander.  The model uses the gas density, velocity vector field, and temperature predicted by computational fluid dynamics (CFD) or direct simulation Monte Carlo (DSMC) simulations to compute the forces and accelerations acting on the regolith particles, one particle at a time (ignoring particle collisions until more advanced models are developed).  Here we present significant improvements to the model, including the implementation of particle drag and lift formulas to account for the rarefaction and compressibility of the flow.  It turns out that the drag force is reduced due to the rarefaction, but the lift is increased due to several effects such as particle rotation.

A data matrix of particle sizes, engine thrusts (descent and ascent values for Altair), horizontal and vertical starting distances, and lander height above ground, have been tested using the latest version of the software.  These results suggest that the previously reported 3 degree trajectory angle limit can be exceeded in several cases by as much as a factor of five. Particles that originate at a height of 1 cm above the surface from an outer crater rim can be propelled to angles of 5 degrees or more.  Particles that start at 10 cm above the surface can be ejected with angles of up to 15 degrees.  Mechanisms responsible for placing particles at starting heights above the surface may include the kinetics of horizontal collisions, as suggested by Discrete Element Method (DEM) simulations.  We also present results showing the distance particles travel and their impact velocities.  We then use the model to evaluate the effectiveness of berms or other methods to block the spray of soil at a lunar landing site.

## INTRODUCTION

Phase II of a mathematical model and software implementation developed to predict trajectories of single lunar dust particles acted on by a high velocity gas flow is discussed.  The model uses output from a computation fluid dynamics (CFD) or direct simulation Monte Carlo (DSMC) simulation of a rocket nozzle hot gas jet.  The gas density, velocity vector field, and temperature predicted by the CFD/DSMC simulations, provide the data necessary to compute the forces and accelerations acting on a single particle of regolith. As in Phase I (Lane 2008), all calculations of trajectory assume that the duration of particle flight is much shorter than the change in gas properties, i.e., the particle trajectory calculations take into account the spatial variation of the gas jet, but not the temporal variation.

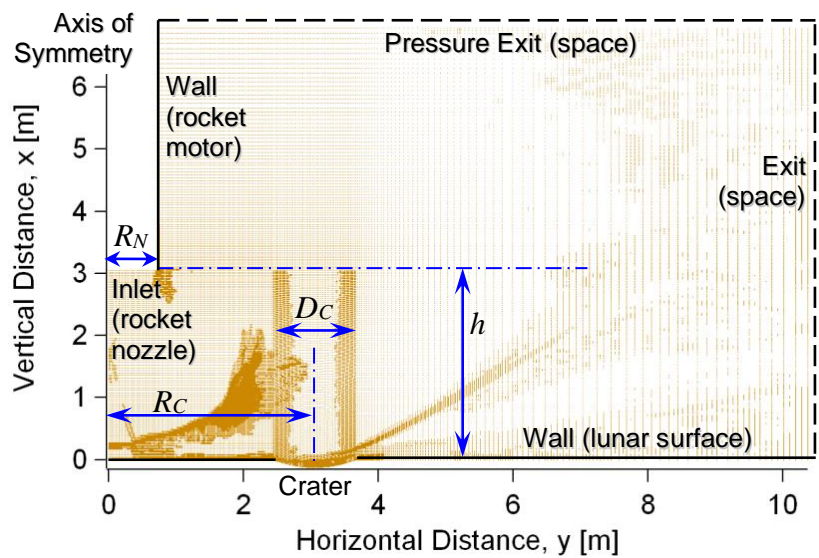

**Figure 1.  2D CFD boundary and grid point definition.**

The trajectory model described in this paper requires output from a two- or three-dimensional fluid dynamics simulation of the rocket nozzle exhaust. The worked described in the Phase I work utilized both CFD and DSMC simulations of an Apollo-like lunar Lander with a maximum descent thrust of approximately 6.3 kN. In the present Phase II work, the gas densities from the Apollo 2D CFD and 3D DSMC simulations are scaled up by a factor of 5.4 (resulting in a thrust of 33.7 kN) representing descent, and by 13.2 for ascent (resulting in a thrust of 82.5 kN) for the Altair lander.

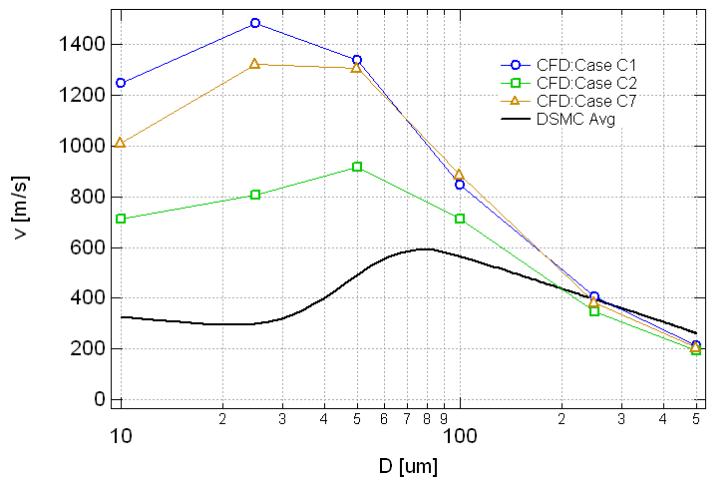

**Figure 2a. Particle speeds exiting the CFD boundary for various particle sizes and CFD Cases of Table 1. The solid black line is averaged data from Metzger (2007).**

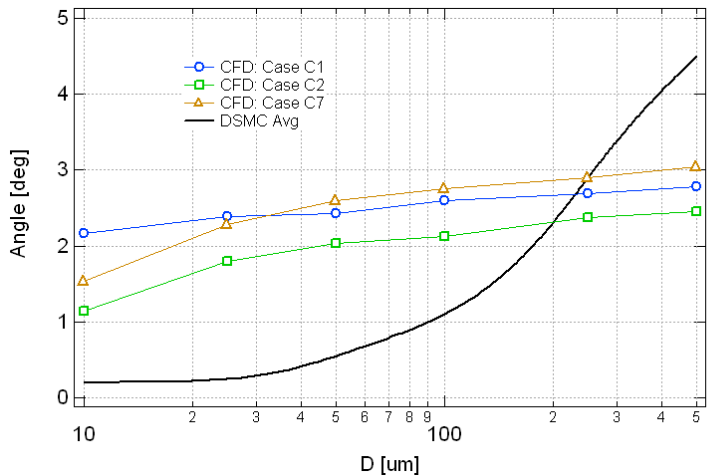

**Figure 2b. Particle trajectory angles relative to ground corresponding to Figure 7. Again, the solid black line is averaged data from Metzger (2007).**

### Background: Phase I Software (2007)

The CFD/DSMC output provides estimates of gas density $\rho(\mathbf{r})$, gas velocity $\mathbf{u}(\mathbf{r})$, and gas temperature $T(\mathbf{r})$ for a gridded volume described by $\mathbf{r}_i$ vector points in the bounded domain. These values are interpolated from the CFD/DSMC grid by finding four nearest grid neighbors in a volume around the *i*th trajectory point and applying an *N*-dimensional interpolation algorithm based on a direct implementation of Shepard's interpolation (Shepard 1968). Inputs which are initial conditions of the trajectory calculation include the particle diameter *D* and initial position of the particle $\mathbf{r}_0 = (x_0, y_0, z_0)$ where the vertical direction *x* is positive up and equal to zero at the surface. Typically, the particle starting position might be resting on the lunar surface at $x = 0$, so that $x_0 = D/2$. Results from the Phase I work as shown in Figures 2 (Lane 2008).

### Phase II Software Improvements (2009)

In Phase II, many software algorithm refinements have been made. Perhaps the most important of these improvements are new formulas for particle drag and lift. These formulas are taken directly from Loth (2008a). Based on these new results, drag is reduced due to Knudsen number effects, but lift is increased due to several effects such as particle rotation. The Loth lift model is described in detail in the references (Loth 2008b).

The interpolation algorithm in the 2007 version of trajectory code was replaced by code developed at the University of North Texas (Renka 1988a, 1988b, 1988c). QSHEP2D is an implementation of the modified quadratic Shepard method for the case of two independent variables. The software conforms to both the 1966 and 1977 ANSI Standards for FORTRAN, and has no system dependencies. Header comments in each routine contain detailed descriptions of the calling sequences, and all parameter names conform to the FORTRAN typing default. QSHEP3D is the 3D version of QSHEP2D and is included in the new version of the trajectory code. QSHEP3D is used to interpolate DSMC input data, whereas QSHEP2D is used to interpolate CFD data.

| Height of Nozzle above Surface, $h$ | Distance from Nozzle Center Line to Crater Center, $R_C$ | |
|---|---|---|
| | $5\,R_N$ | $15\,R_N$ |
| $2.5\,R_N$ | Case C2 | --- |
| $5\,R_N$ | Case C7 | Case C3 |
| $10\,R_N$ | Case C1 | --- |

**Table 1. CFD cases used as inputs to trajectory model.**

## SIMULATION EXPERIMENTS

Three CFD cases have been used as the input to the current software version PTQv54: cases C1, C7, and C2. The particle trajectories due to these three cases have been studied extensively. Each case represents a different Altair lander height above the lunar surface during descent, starting at h = 20 ft, then 10 ft, and 5 ft. The simulation data discussed contains complete trajectory simulation results for 11 horizontal particle starting points and two vertical locations, for a total of 22 initial starting points. In addition, five particle sizes, from 1 μm to 10000 μm, in multiples of 10 are part of the matrix. This complete matrix set corresponds to approximately 100 hrs of CPU time.

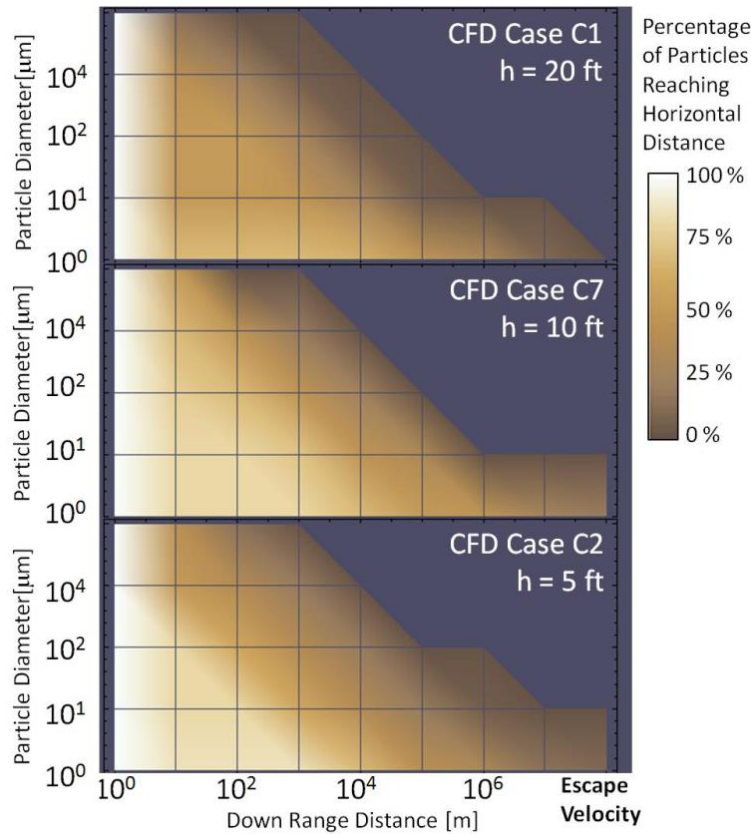

**Figure 3. Plot color corresponds to the percentage per diameter bin of particles that make it to or past the down range value on the horizontal axis.**

Referring to Figure 3, the horizontal axis is the down range distance (in m) of the trajectory; vertical axis is particle diameter (in μm); and plot color corresponds to the percentage per diameter bin of particles that make it to or past the down range value on the horizontal axis (similar to a cumulative PDF). The simulation particles consisted of 11 horizontal starting points (1 to 7 m) and 2 vertical starting points (0.01 and 0.1 m). These cases correspond to the proposed Altair lunar lander descent thrust equal to approximately 33.7 kN, a factor of 5.4 greater than the Apollo lander thrust. Note that escape velocity, $v_E = \sqrt{2 g_L R_L} \approx 2.37$ km/s, where $g_L =$ 1.622 m/s² and $R_L$=1737 km, is reached in some cases of the smallest particles.

The following two pages show 3D Mathematica plots of velocity and angle as a function of particle diameter and radial starting point (Figures 4). The velocity and angle correspond to the particle trajectory as it leaves the CFD domain at approximately 10 m. Note that some of the small particles manage to exceed escape velocity. There are two sets of plots for the three lander heights, where each set consists of two particle starting heights, $y_0$, equal to 0.01 m and 0.10 m. Note that in all cases, the 4 ft diameter crater centered at 10 ft, has a dramatic effect on the particle trajectories. Most notable is the higher trajectory angle when the particle starts at the outer crater rim at a radial distance of 12 ft.

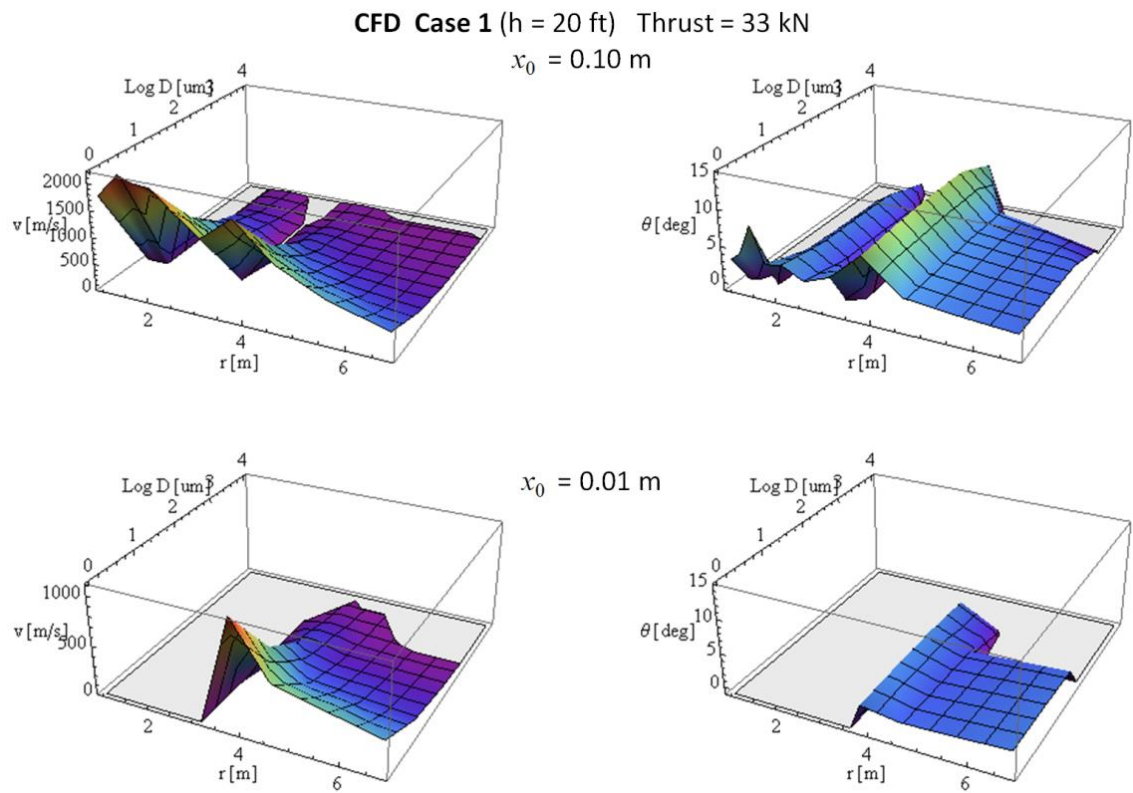

**Figure 4a.**

**CFD Case 7** (h = 10 ft)   Thrust = 33 kN
$x_0 = 0.10$ m

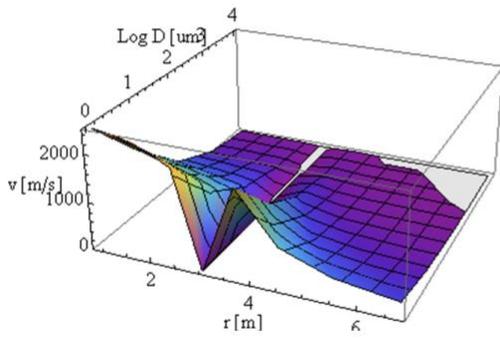
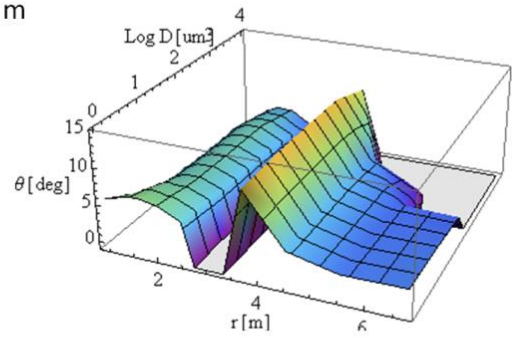

$x_0 = 0.01$ m

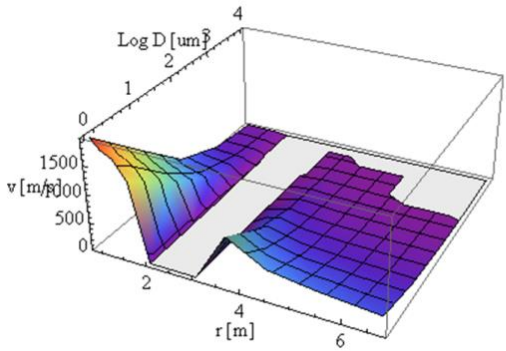

**Figure 4b.**

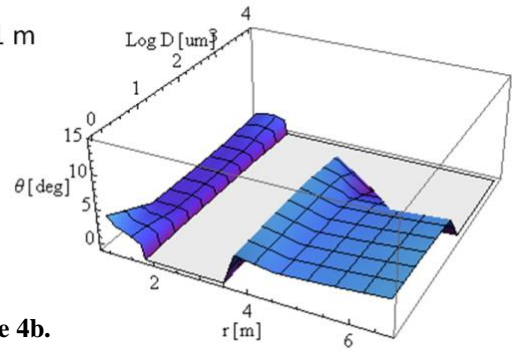

**CFD Case 2** (h = 5 ft)   Thrust = 33 kN
$x_0 = 0.10$ m

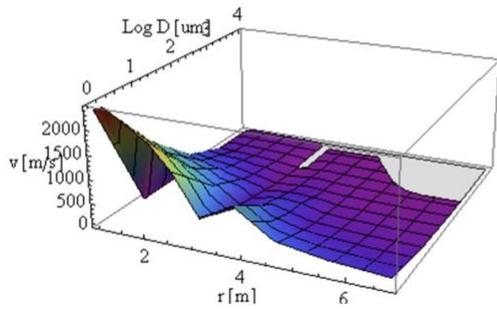
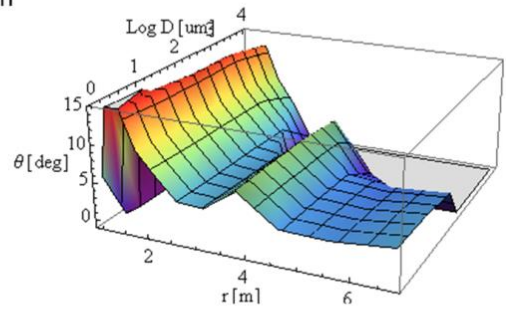

$x_0 = 0.01$ m

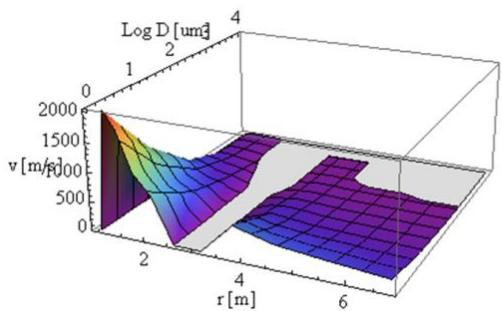

**Figure 4c.**

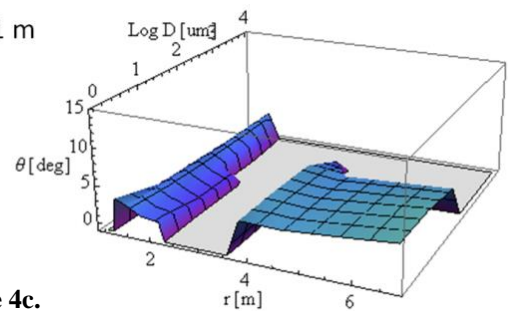

## High Angle Particle Ejection

A new simulation effort was undertaken using version PTQv55 of the particle trajectory modeling code to help answer the question: *how far away should the landers land from the lunar outpost?* The simulation focus is on Altair landing thrust levels (~ 34 kN) with a 2 m high berm at 30 m. For purposes of modeling, the target (outpost structure) is set at 1 km and is 3 m tall. Based on the simulations, the berm does not stop all particles. The reason for this is evident in the previous section. By examining the CFD case 2 ejection angle plot between 1 and 2 m horizontal distance, it can be seen that particles of all diameters simulated (1 – 10000 μm) are ejected at high angles, greater than 10 deg. It is difficult to extract information predicting the percentage of particles stopped by the berm and the percentage not stopped, since there is no particle size distribution information in the simulations. The best guess for percentages predictions is based on treating the circular surface under the lander as containing uniform size distributions. The percentage stopped by a berm in the simulation is specified as a function of particle size. The total number of particles could then be estimated by multiplying the simulated percentage of particles not stopped (or stopped) by the nominal particle size distribution under the lander.

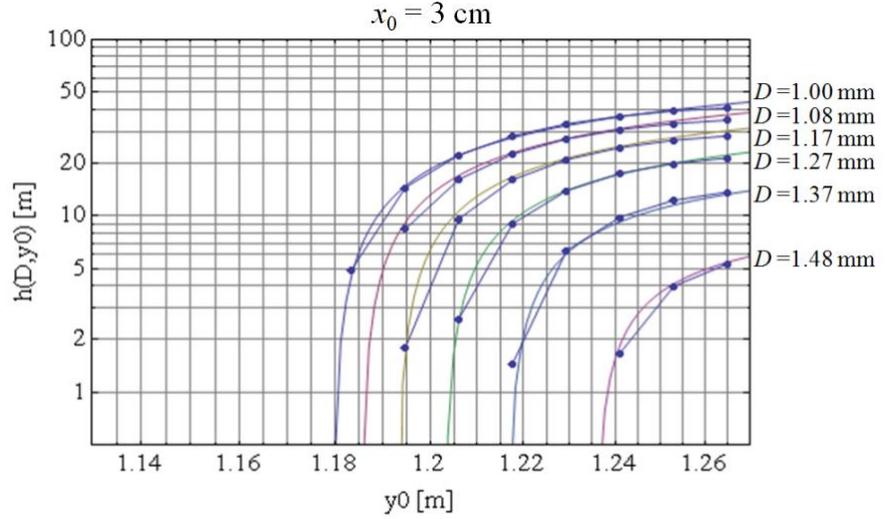

**Figure 5. An example of the curve fitting obtained using Equations (1) and (2) is shown in Figure 5 for the case of $x_0 = 3$ cm. Filled circles are data from original simulations.**

Since this is a special case and a particularly interesting one, an empirical multi-parameter curve fit was applied to the previous simulation. By plotting the height $h$ that a particle passes over the target set at 1 km, a good empirical fit is found by:

$$h(D, y, x) = a \ln \left( b(y - y_0) \right) \qquad (1)$$

where,

$$a(D) = \exp\left( \alpha_0 + \alpha_1 \exp(\alpha_2 D) \right) \qquad (2a)$$

$$b(D) = \exp\left( \beta_0 + \beta_1 \exp(\beta_2 D) \right) \qquad (2b)$$

$$y_0(D) = \exp\left( \gamma_0 + \gamma_1 \exp(\gamma_2 D) \right) \qquad (2c)$$

An example of the original simulation from the previous section and the curve fitting obtained using Equations (1) and (2) is shown in Figure 5 for the case of $x_0 = 3$ cm.

Table 2 shows the results of the curve fit for five starting heights from 3 to 7 cm in 1 cm increments.

A pseudo particle distribution can be generated from Equation (1) by scanning over a small local region around $y$ and $x$ and by applying the following test:

**Table 2. Coefficients from Equations (1) and (2) fitted to simulations results.**

| $x$ [cm] | $j$ | $\alpha_j(x)$ | $\beta_j(x)$ | $\gamma_j(x)$ |
|---|---|---|---|---|
| 3 | 0 | 3.641 | 3.679 | 0.1112 |
|   | 1 | -0.01001 | 0.01079 | 0.005455 |
|   | 2 | 3.705 | 3.219 | 1.925 |
| 4 | 0 | 3.938 | 1.302 | 0.06165 |
|   | 1 | -0.005320 | 1.3934 | 0.02681 |
|   | 2 | 3.242 | 0.4878 | 0.8397 |
| 5 | 0 | 4.765 | 3.062 | 0.03930 |
|   | 1 | -0.09739 | 0.07943 | 0.03693 |
|   | 2 | 1.473 | 1.507 | 0.6007 |
| 6 | 0 | 4.328 | 3.85237122 | 0.08385 |
|   | 1 | -0.005461 | $3.205 \times 10^7$ | 0.008680 |
|   | 2 | 2.291 | 5.858 | 0.8994 |
| 7 | 0 | 4.476 | 3.768 | 0.06055 |
|   | 1 | -0.005787 | 0.0002876 | 0.01778 |
|   | 2 | 2.038 | 2.917 | 0.6189 |

$$n(D, y, x) = \begin{cases} n(D, y, x) + 1 & h(D, y, x) > 0 \ \& \ h(D, y, x) < h_0 \\ no\ change & otherwise \end{cases} \quad (3)$$

where $h_0$ is the height of the target, set to 3 m for this example. Once the size distribution is determined, several useful quantities can be calculated. For example, the average size particle hitting the target can be computed as the *first moment of the distribution*:

$$\bar{D}(y, x) = \frac{\int_0^\infty D\, n(D, y, x)\, dD}{\int_0^\infty n(D, y, x)\, dD} \quad (4)$$

The standard deviation can be computed by the *second moment*:

$$\sigma_D^2(y, x) = \frac{\int_0^\infty (D - \bar{D})^2\, n(D, y, x)\, dD}{\int_0^\infty n(D, y, x)\, dD} \quad (4)$$

Using Equations (4) and (5), the origin of average diameter particle and its standard deviation striking the target (3 m high, at 1 km) can be plotted, as shown in the following Figures 6.

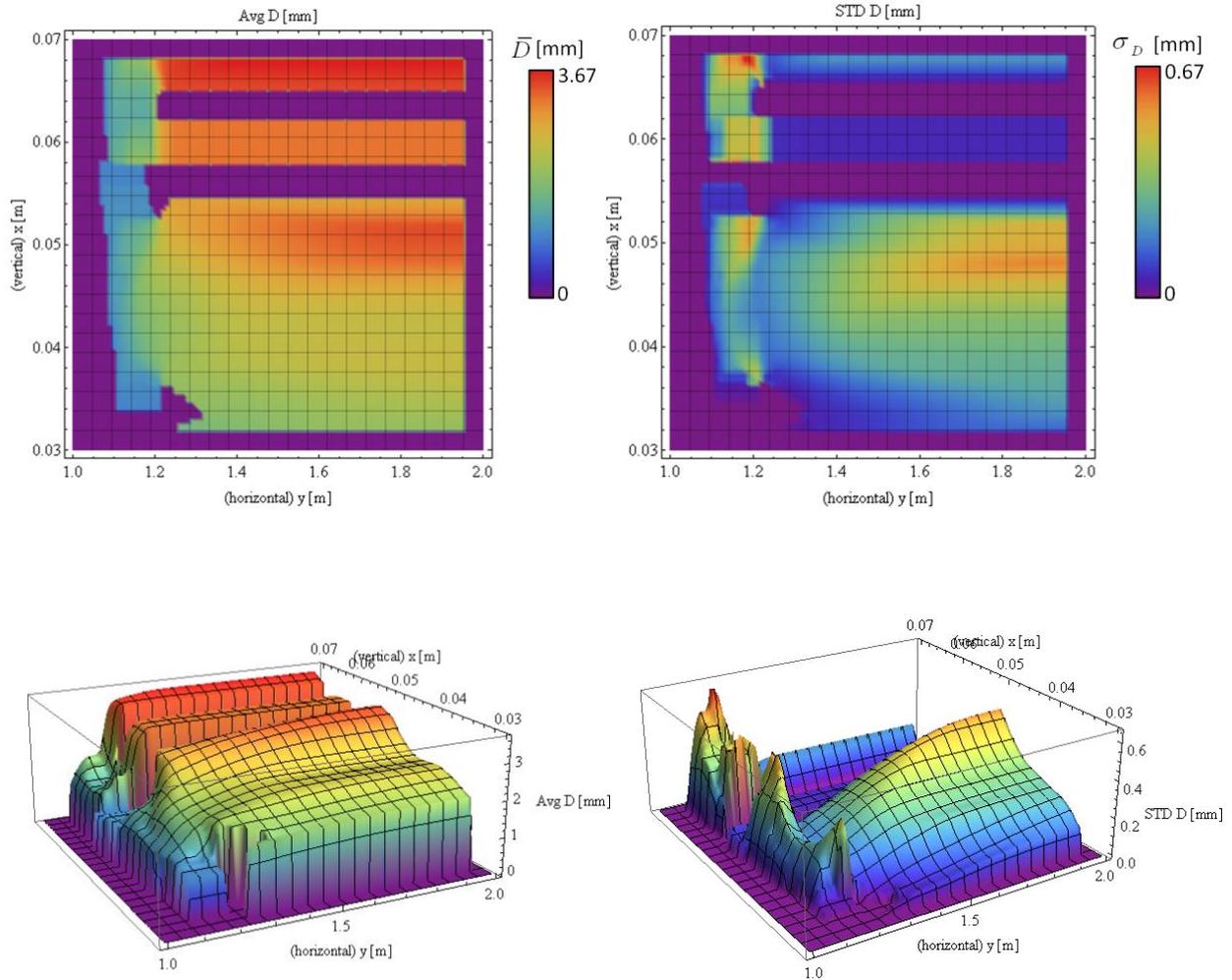

**Figure 6. Particle diameter and location of average size (left) and standard deviation (right) striking a target, 3 m high, at 1 km.**

**SUMMARY**

The results of the particle trajectory simulations strongly suggest that particles up to 1 cm in diameter will under certain conditions be ejected at higher angles than what has previously been believed. Previous work showed most particles ejected from a rocket jet impingement on the lunar surface are in the range of 3 deg or less. A 2 m high berm located at 30 m should be effective in stopping particles with trajectory angles of up to 4 deg. Based on these simulations, some larger particles will be ejected at angles of up to 15 deg near the rocket nozzle. These larger particles have relatively low velocities, but due to the lack of any drag impediment, these particles can travel long distances before raining down to the ground.

The next step in these simulations is to incorporate an estimate of the actual particle size distribution that is lifted into the jet stream. From that, the number of particles and flux impinging an area broadside can be computed. Obviously, there will be a very large reduction in the number of particles hitting an outpost structure located at 1 km due to spherical spreading. These results so far have been computed from the CFD simulations of 2007. To verify validity of the predictions, the DSMC simulations from 2007 can also be used as input to the particle trajectory model. This is a planned activity for future work.